\documentclass[prl,twocolumn,showpacs]{revtex4}
\usepackage[dvips]{graphicx}
\usepackage{latexsym}
\usepackage{amsmath}
\usepackage{amsfonts}
\usepackage{amssymb}
\usepackage{bm}
\usepackage{color}
\begin{document}
\newcommand{\fig}[2]{\includegraphics[width=#1]{#2}}
\newcommand{{\vhf}}{\chi^\text{v}_f}
\newcommand{{\vhd}}{\chi^\text{v}_d}
\newcommand{{\vpd}}{\Delta^\text{v}_d}
\newcommand{{\ved}}{\epsilon^\text{v}_d}
\newcommand{{\vved}}{\varepsilon^\text{v}_d}
\newcommand{{\bk}}{{\bf k}}
\newcommand{{\bq}}{{\bf q}}
\newcommand{{\tr}}{{\rm tr}}
\newcommand{\pprl}{Phys. Rev. Lett. \ }
\newcommand{\pprb}{Phys. Rev. {B}}

\title{Chiral spin density wave order on frustrated honeycomb and bilayer triangle lattice Hubbard model at half-filling}

\author{Kun Jiang,$^1$ Yi Zhang,$^1$ Sen Zhou,$^{2}$ and Ziqiang Wang$^1$}
\affiliation{$^1$ Department of Physics, Boston College, Chestnut Hill, MA 02467,
USA}
\affiliation{$^2$ State Key Laboratory of  Theoretical Physics, Institute of
Theoretical Physics, Chinese Academy of Sciences, Beijing 100190, China}
\date{\today}

\begin{abstract}
We study the Hubbard model on the frustrated honeycomb lattice with nearest-neighbor $t_1$ and second nearest-neighbor hopping $t_2$, which is isomorphic to the bilayer triangle lattice, using the SU(2)-invariant slave boson theory. We show that the Coulomb interaction $U$ induces antiferromagnetic (AF) chiral spin-density wave ($\chi$-SDW) order in a wide range of $\kappa =t_2/t_1$ where both the two-sublattice AF order at small $\kappa$ and the decoupled three-sublattice 120$^\circ$ order at large $\kappa $ are strongly frustrated, leading to three distinct phases with different anomalous Hall responses. We find a continuous transition from a $\chi$-SDW semimetal with anomalous Hall effect to a topological chiral Chern insulator exhibiting quantum anomalous Hall effect, followed by a discontinuous transition to a $\chi$-SDW insulator with zero total Chern number but anomalous ac Hall effect.
The $\chi$-SDW is likely a generic phase of strongly correlated and highly frustrated hexagonal lattice electrons.

\typeout{polish abstract}
\end{abstract}


\maketitle

A spin density wave (SDW) refers to the formation of nonzero spin density moments in itinerant electron systems \cite{overhauser}. The spin texture depends on the nature of the electronic interaction, the lattice geometry and the Fermi surface (FS) structure. It has the general form:
$\vec S(\vec r)=\sum_{\alpha}\vec S_{\alpha}\cos(\vec Q_\alpha\cdot \vec r+\theta_\alpha)$
where $\alpha=x,y,z$ and $\theta_\alpha$ is a relative phase. The SDW ordering wavevectors $\vec Q_\alpha$, when commensurate with the lattice, determine the magnetic unit cell containing a number of sublattice sites. Besides the usual linearly-polarized (collinear) and spiral (coplanar) SDW phases, the textured quantum electronic phase with noncoplanar, chiral SDW ($\chi$-SDW) order has attracted great interest recently for its ability to sustain a spin chirality $\chi=\vec S_{\ell_1}\cdot (\vec S_{\ell_2}\times \vec S_{\ell_3})$, where $\ell_i$ labels the sublattice sites in the magnetic cell, that breaks both parity and time-reversal symmetry. Electrons accumulate Berry phase from the spontaneous internal magnetic field, leading to the anomalous Hall effect (AHE) \cite{luttinger,ye99,nagaosa00,nagaosa01a,nagaosa01b}. A topological phase with quantum anomalous Hall effect (QAHE) can arise in a $\chi$-SDW insulator, where the electron bands acquire nonzero Chern numbers \cite{nagaosa00}.

The spin-chirality mechanism accounts for the AHE in many ferromagnetic materials such as the manganites and the pyrochlores \cite{nagaosa10}. In this paper, we focus on the antiferromagnetic (AF) $\chi$-SDW metals and insulators with $\sum_\ell \vec S_\ell=0$ in materials and models with strong electron correlation and magnetic frustration. They have been discovered in charge transfer insulators NiS$_2$ \cite{miyadai75,kikuchi78a,kikuch78b,matsuura03}, metallic $\gamma$-FeMn alloys \cite{endoh71,tajima76,kennedy87,kawarazaki90} and related materials where the magnetic moments reside on the frustrated face-centered-cubic lattice. Neutron scattering observed noncoplanar AF order with 4-sublattices and 3-ordering wavevectors. A unique character of this triple-$Q$ $\chi$-SDW phase is that the ordered moments on the four sublattices form a tetrahedron in spin space. On the theoretical side, it has been shown that frustrated Heisenberg two-spin exchange interactions are insufficient to produce the AF $\chi$-SDW order; additional 4-spin exchange interactions are necessary for such a noncoplanar SDW to emerge from the many degenerate magnetic states \cite{yoshida81,yoshimori81,kunitomo85,matsuura03}. In addition, weak-coupling approaches such as nesting based models \cite{maki76} and band structure (LDA) and LDA+U calculations \cite{sakuma00,matsuura03,ekholm11} have been performed to study the complex magnetic order in these materials. While a microscopic theory for the $\chi$-SDW order is currently lacking, it is believed that both strong correlation and geometric frustration play vital roles in its origin.

Recently, it has been shown that the nearest neighbor (NN) Hubbard model on the triangular and the honeycomb lattices has a FS instability at $3/4$ zone-filling, where the FS touches the van Hove (vH) singularity \cite{martin08,taoli12,hayami14}. In the magnetic channel, this instability leads to the same triple-$Q$ $\chi$-SDW order with 4-sublattice spins forming a tetrahedron. Several theoretical studies such as the renormalization group (RG) \cite{chubukov12}, functional RG \cite{wang12,kiesel12}, and density matrix RG \cite{ran14} have been performed to study the competition of the $\chi$-SDW state with other forms of FS instabilities such as unconventional superconductivity. These findings raise the exciting possibility of realizing the AHE and the topological QAHE in {\em two-dimensional} (2D) or layered quasi-2D materials such as graphene \cite{graphenea,grapheneb}, sodium cobaltates \cite{takada03,singh03,pickett04}, and frustrated antiferromagnets \cite{lee05,shiomi12}, and motivate the study of topological AF $\chi$-SDW ground states in 2D models with electronic correlation and geometric frustration.

    \begin{figure}
     \begin{center}
    \fig{3.2in}{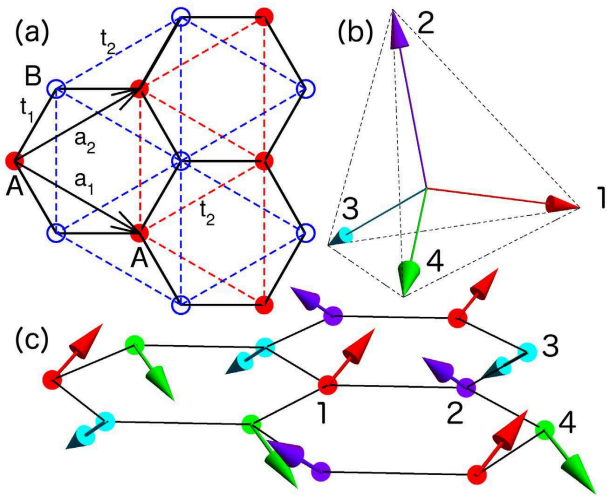}\caption{(a) The isomorphic $t_1$-$t_2$ honeycomb lattice and bilayer triangle lattice. (b) and (c) Spin configurations of the 4-site, triple-$Q$, tetrahedron AF $\chi$-SDW order.}
    \end{center}
    \vskip-0.5cm
    \end{figure}

We study in this work the frustrated honeycomb lattice Hubbard model with NN $t_1$ and second NN hopping $t_2$, and on-site Coulomb repulsion $U$ shown in Fig.~1a. This lattice structure is isomorphic to the center-stacked, bilayer triangle lattice with intralayer hopping $t_2$ and interlayer hopping $t_1$ as indicated by the dashed blue and red lines in Fig.~1a. Materials having such lattice structures include, in addition to graphene and bilayer cobaltates, quasi-2D bilayer triangular lattice chalcogenides \cite{nakatsuji07} and layered honeycomb lattice AF compounds \cite{shiomi12,matsuda10,singh10,li14}. We study the model at {\em half filling} in view of the better control over stoichiometric materials, and address the nature of the magnetic ground states at large enough $U$ that straddle between the two-sublattice collinear AF order at $t_2/t_1\ll 1$ and the two decoupled 120$^\circ$ coplanar order at $t_1/t_2\ll 1$.
To study both strong correlation and noncollinear magnetic order, we employ the SU(2) spin rotation invariant slave boson theory \cite{kr86,li89,wolfle92}, which has been generalized to treat magnetic superstructures \cite{jiang14}. This approach describes the magnetism on the square lattice that shows remarkable agreement with QMC simulations \cite{lilly90}. Recently, the semimetal to AF insulator transition on the honeycomb lattice was studied using this approach \cite{zhou14} and the obtained results agree well with the QMC work \cite{sorella12}. On the frustrated triangular lattice, the SU(2)-invariant slave boson theory predicts a discontinuous transition to the noncollinear 120$^\circ$ AF ordered phase at a critical $U$ \cite{jiang14} that is also in good agreement with variational Monte Carlo \cite{wantanabe08} and numerical RG calculations \cite{yoshioka09}. Fig.~2 shows our obtained phase diagram on the axes of frustration $t_2/t_1$ and correlation $U/W$, where $W$ is the bandwidth. We find that in a wide range of $t_2/t_1$ where the AF frustration is most pronounced, the ground state for $U>0.68W$ is precisely the triple-$Q$, noncoplanar $\chi$ SDW phase 
shown in Figs.~1b and 1c with ordering wavevectors $\vec Q_{1,2}  = {1\over2}\vec b_{1,2}$ and $\vec Q_3={1\over2}(\vec b_1+\vec b_2)$, where $\vec b_{1,2}$ are the reciprocal lattice vectors of $\vec a_{1,2}$ in Fig.~1a. Furthermore, Fig.~2 shows several distinct and novel SDW phases protruding into the paramagnetic (PM) phase at smaller $U/W$, forming the triangle-shaped phase region around $t_2^*=0.85t_1$. Increasing the Hubbard $U$ drives a sequence of phase transitions: from the PM metal to a single-$Q$ striped SDW, then to a double-$Q$ coplanar spiral SDW, followed by the onset of triple-$Q$ $\chi$-SDW order into a semimetal with AHE, then to a topological $\chi$-SDW exhibiting QAHE with the total occupied-band Chern number $C=2$; and eventually via a discontinuous topological transition to a $\chi$-SDW insulator with $C=0$ and spontaneous ac AHE.
%

    \begin{figure}
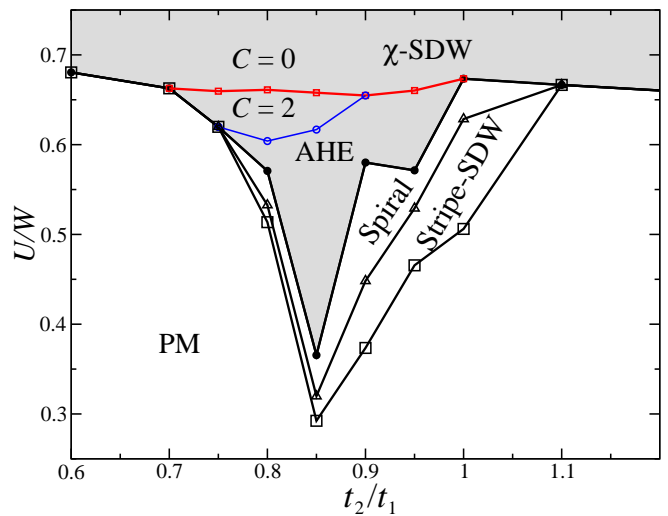

      \begin{center}
    \fig{3.4in}{pd.eps}\caption{Phase diagram: Shaded regions represent distinct $\chi$-SDW phases: Semi-metal with AHE, $C=2$ Chern insulator with QAHE, and $C=0$ Chern insulator with ac AHE. All phase boundaries are lines of discontinuous transitions except for the continuous transition across the blue line.}
    \end{center}
    \vskip-0.5cm
    \end{figure}

The Hubbard model on the lattice shown in Fig.~1a is
\begin{equation}
H = \sum_{\langle ij \rangle } {t_1 } c_{i\sigma }^ \dagger  c_{j\sigma }  + \sum_{\langle\langle ij \rangle\rangle} {t_2 } c_{i\sigma }^\dagger  c_{j\sigma }  + {\rm h.c.} + U\sum_i {n_{i \uparrow } n_{i \downarrow } },
\end{equation}
where $t_1$ and $t_2$ describe the inter-sublattice (inter-layer) and intra-sublattice (intralayer) hopping on the honeycomb (bilayer triangle) lattice. Labeling the two-sublattice (bilayer) as $A$ and $B$ and denoting $C_{k\sigma}^\dagger= (c_{k\sigma,A}^\dagger, c_{k\sigma,B}^\dagger)$, the noninteracting part in Eq.~(1) can be written as \cite{lin09} $H_0=C_{k\sigma}^\dagger H_k C_{k\sigma}$, where
\begin{equation}
 H_k = \left( {\begin{array}{*{20}c}
   {t_2 \Delta_k} & {t_1\varepsilon_k }  \\
   {t_1\varepsilon_k^* } & {t_2 \Delta_k}  \\
\end{array}} \right), \quad \varepsilon_k  = 1 + e^{ - i\vec k \cdot \vec a _1 }  + e^{ - i\vec k \cdot \vec a _2 },
\end{equation}
and $\Delta_k = 2[\cos(\vec k \cdot \vec a _1 ) + \cos(\vec k \cdot \vec a _2 ) + \cos(\vec k \cdot (\vec a _1  - \vec a _2 ))]$. Diagonalizing $H_k$ gives two noninteracting bands
$
E_k^\pm  = t_2 \Delta_k \pm t_1 \sqrt {3 + \Delta_k}.
$
For $t_2<t_1/3$, the two subbands cross at the Dirac points ($K$ and $K^\prime$) that pin the Fermi level at {\em half-filling}. When $t_2>t_1/3$, the subbands overlap, giving rise to three FS sections: a hole pocket around zone center ($\Gamma$) and two electron pockets around $K$ and $K^\prime$ as shown in Fig.~3(a). Increasing $t_2/t_1$ further, the FS pockets grow in size and the Fermi level rises toward the vH singularity at $M$ point with energy $E_{M}^+=t_1-2t_2$. Fig.~3(b) shows that at {\em half-filling}, the Fermi level touches the vH points at $t_2^*\simeq0.85t_1$ where the electron pockets merge and the hole pocket matches the $2\times2$ reduced zone boundary.
For $t_2>t_2^*$, the electron pockets coalesce to produce the large hexagonal electron FS (Fig.~3c), while the central hole pocket grows continuously.
A weak-coupling theory of itinerant electrons would thus predict an SDW instability at $t_2/t_1\simeq0.85$, associated with both the hexagonal electron FS due to the vH singularity and the hole FS due to umklapp scattering, involving some or all of the three relevant wavevectors $\vec Q_{1,2,3}$ shown in Fig.~3(a).

    \begin{figure}
    \begin{center}
    \fig{3.4in}{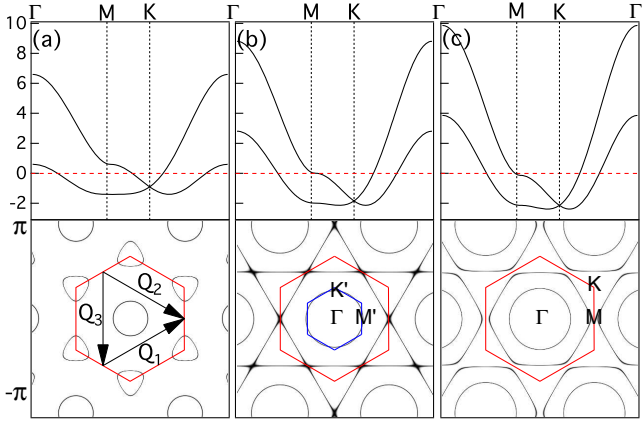}\caption{Band dispersion (top panel, in unit of $t_1$) and FS (bottom panel) at $t_2/t_1=0.5$ (a); $t_2/t_1=0.85$ (b); and $t_2/t_1=1.0$ (c). Red hexagons in the bottom panel mark the original zone boundary, on which lie six vH points connected by the wave vectors $\vec Q_{1,2,3}$ shown in (a). The outer FS crosses the vH points in (b) where the blue hexagon indicates the $2\times2$ reduced zone boundary that intersects the inner hole-FS.}
    \end{center}
    \vskip-0.5cm
    \end{figure}

To treat Coulomb interaction nonperturbatively and study noncollinear spin order, we represent the local Hilbert space
by a spin-1/2 fermion $f_\sigma$ and six bosons $e$, $d$, and $p_\mu$ ($\mu=0,1,2,3$) for empty, doubly-occupied, and singly occupied sites respectively \cite{kr86,li89,wolfle92,jiang14}: $\vert0\rangle=e^\dagger \vert\text{vac}\rangle$,  $|\!\!\uparrow\downarrow\rangle=d^\dagger f_\downarrow^\dagger f_\uparrow^\dagger \vert\text{vac}\rangle$, and $\vert \sigma\rangle= {1\over\sqrt{2}} f_{\sigma^\prime}^\dagger p_\mu^\dagger \tau_{\sigma^\prime\sigma}^\mu \vert \text{vac}\rangle$ where ${\tau}^{1,2,3}$ and ${\tau}^0$ are Pauli and identity matrices.
The completeness of the Hilbert space, and the equivalence between boson and fermion representations of the particle and spin density impose three local constraints:
\begin{eqnarray}
 O_i&=&e_i^\dagger  e_i + p_{i0}^\dagger  p_{i0}  + \vec p_i^\dagger   \cdot \vec p_i  + d_i^\dagger  d_i -1=0, \nonumber \\
 O_{i}^0&=&p_{i0}^\dagger  p_{i0}  + \vec p_i^\dagger   \cdot \vec p_i  + d_i^\dagger d_i - f_{i\sigma}^\dagger  f_{i\sigma}  = 0, \nonumber \\
 O_{i}^\alpha&=& p_{i0}^\dagger  p_{i\alpha}  + p_{i\alpha}^\dagger p_{i0} + i(\vec p_i^\dagger\times \vec p_i)_\alpha - f_{i\sigma}^\dagger\tau _{\sigma\sigma^\prime}^\alpha f_{i\sigma^\prime}  = 0.
 \nonumber
 \end{eqnarray}
The Hubbard Hamiltonian thus becomes,
\begin{eqnarray}
H &= & \sum_{\langle ij \rangle} t_1 \psi_{i}^\dagger  g_i^\dagger g_j \psi_j
  +\sum_{\langle\langle ij \rangle\rangle} t_2 \psi_{i}^\dagger  g_i^\dagger g_j \psi_j + U\sum_i {d_i^\dagger  d_i } \nonumber \\
    &-&\mu _0 \sum\limits_i f_{i\sigma }^\dagger  f_{i\sigma } + \sum_i\lambda _i O_i  + \sum_i \lambda _{i\mu} O_{i}^\mu,
\label{slavebosonh}
\end{eqnarray}
where the fermion spinor $\psi_i^\dagger=(f_{i\uparrow}^\dagger, f_{i\downarrow}^\dagger)$ and $\lambda_i$ and $\lambda_i^\mu$ are Lagrange multipliers. The hopping renormalization factors $g_i$, $g_j$ are $2\times2$ matrices involving the boson operators \cite{li89,wolfle92}. We found that due to the particle-hole symmetry at half-filling, $g_i$ simplifies considerably when all bosons are condensed and $g_i=g_{i0}\tau_0$ where $g_{i0}$ is the corresponding hopping renormalization of Kotliar and Ruckenstein \cite{kr86}. We solve the self-consistency equations that minimize Eq.~(\ref{slavebosonh}) for general spin and charge configurations containing up to $8$-sites per unit cell. To determine the ground state properties accurately, we use the supercell construction \cite{jiang14} and discretize the reduced zone with $600\times600$ $k$-points in all calculations such that uncertainties are within the symbol sizes in Fig.~2.

    \begin{figure}
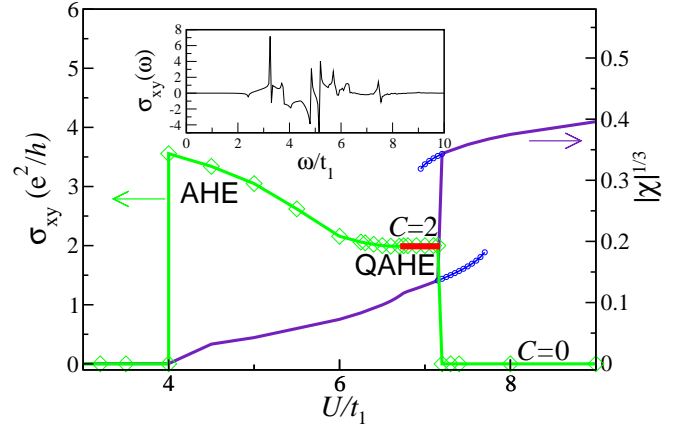

      \begin{center}
    \fig{3.4in}{ps.eps}\caption{Evolution of spin chirality and anomalous Hall response as a function of $U/t_1$ at $t_2/t_1=0.85$. Calculated dc AHE response corresponds to green line, while $C=2$ QAHE is marked by superposed red line. Hysteretic spin chirality (blue line) is shown at transition between $C=2$ and $C=0$ Chern insulators.  Inset: ac AHE response at $U/t_1=8.0$.}
    \end{center}
    \vskip-0.5cm
    \end{figure}

The obtained results show that for $t_2/t_1<0.55$, the bipartite collinear AF insulator remains the ground state as in the unfrustrated case at $t_2=0$ and $U/W \ge0.57$. In the opposite limit, when $t_2/t_1 > 1.3$, the $120^\circ$ coplanar AF state becomes the ground state, which is analytically connected to the decoupled $120^\circ$ states in the limit $t_1\to0$ and $U/W \ge 1.42$. Remarkably, we find that in the wide region $0.55 <t_2/t_1 < 1.3$ the effects of frustration and vH singularity give rise to three new SDW phases as shown in the phase diagram (Fig.~2). They are described by
\begin{eqnarray}
{\rm 1Q \ Stripe}, \quad {\vec S(\vec r_i)}&=&m(\pm e^{i\vec Q_1\cdot \vec r_i},0,0),
\nonumber \\
{\rm 2Q \ Spiral}, \quad  {\vec S(\vec r_i)}&=&{m\over\sqrt{2}}(\pm e^{i\vec Q_1\cdot \vec r_i}, \pm e^{i\vec Q_2\cdot \vec r_i},0),
\label{sdwforms} \\
{\rm 3Q \  \chi\!- \! SDW}, \quad
{\vec S(\vec r_i)}&=&{m\over\sqrt{3}}(\pm e^{i\vec Q_1\cdot \vec r_i},\pm e^{i\vec Q_2\cdot \vec r_i},e^{i\vec Q_3\cdot \vec r_i}),
\nonumber
\end{eqnarray}
with $\pm$ for $i \in A,B$ respectively and $\vec Q_{1,2,3}$ depicted in Fig.~3a.
Let us fix the degree of frustration at $t_2/t_1=0.85$ and increase the correlation strength $U/W$. Fig.~2 shows that the PM metal undergoes two sequential discontinuous transitions to the 1$Q$-strip and then the 2$Q$-spiral phases. These phases are metallic due to the partial gapping of the FS and break the $C_3$ symmetry.

Increasing $U/W$ further leads to the onset of the triple-Q $\chi$-SDW order through a discontinuous transition. We find that a non-zero spin chirality $\chi$ alone is insufficient to specify the ground state and there are three distinct $\chi$-SDW phases characterizable by their intrinsic Hall responses \cite{luttinger}. The latter can be calculated using the Kubo formula \cite{nagaosa01b,thouless82,niu10,yao04},
\begin{equation}
\sigma_{xy} (\omega) = {e^2\over\hbar}\sum_{k,n \ne m} \frac{[f(\varepsilon_{kn})  - f(\varepsilon_{km})] \mathop{\rm Im} v_x^{nm} v_y^{mn}}{(\varepsilon _{kn}  - \varepsilon _{km} )^2  - (\omega  + i\delta )^2 },
\label{sigmaxy}
\end{equation}
where $\varepsilon_{kn}$ is the dispersion of the $n$th band $\vert nk\rangle$ in the self-consistent solution of Eq.~(\ref{slavebosonh}) as shown in Fig.~5; $f(x)$ is the Fermi function; and $v_{x(y)}^{mn}=\langle km\vert \hat v_{x(y)}\vert kn\rangle$ is the matrix element of the velocity operator. In Fig.~4, we plot the calculated anomalous Hall response and the spin chirality $\chi$ as a function of $U/t_1$. As the system enters the $\chi$-SDW phase, the triple-$Q$ order parameter gaps out the vH points of the outer electron FS in Fig.~3b while the inner hole FS is truncated into small electron and hole pockets by the $2\times2$ reduced zone boundary due to umklapp scattering. This $\chi$-SDW-I semimetal phase exhibits (unquantized) dc AHE as shown in Fig.~4. As the ordered moment grows with increasing $U$, the FS pockets shrink and disappear when the system makes a {\em continuous} transition into the insulating $\chi$-SDW-II phase (Fig.~2) characterized by a QAHE with $\sigma_{xy}=C e^2/h$ and $C=2$ as can be seen in Fig.~4. This topological phase is a Chern insulator (CI), since all bands acquire a nonzero Chern number \cite{fukui05} and the total Chern number of all occupied bands is $C=2$ as displayed in Fig.~5a.

This topological phase should remain stable unless the insulating single-particle gap closes, such as when crossing the sample edges where gapless surface states must emerge. Quite surprisingly, Fig.~4 shows that the $\sigma_{xy}$-plateau collapses above $U=7.2t_1$ where a third $\chi$-SDW-III insulating phase sets in and remains the ground state in the large-$U$ region of the phase diagram, which was explicitly verified up to $U/W=8$. In this phase, each band still carries a nonzero Chern number, as shown in Fig.~5b, but the total occupied band Chern number $C=0$ leading to $\sigma_{xy}=0$. We find that the CIs with $C=2$ and $C=0$ are separated by a discontinuous topological transition as detailed in Fig.~4, which is accompanied a hysteretic jump in the spin chirality $\chi$ {\em without} symmetry change in the order parameter or gap-closing. Although the dc Hall response is zero in the $C=0$ CI phase, the Chern bands shown in Fig.~5b give rise to an intrinsic AHE in the ac Hall response $\sigma_{xy}(\omega)$, related to nontrivial optical dichroism through interband transitions as shown in the inset of Fig.~4.

   \begin{figure}
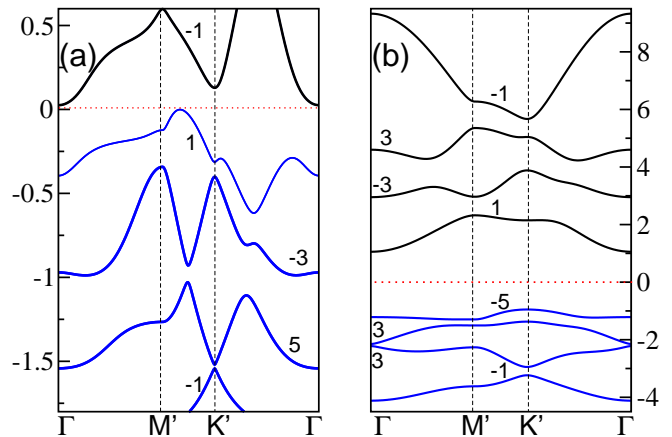

      \begin{center}
    \fig{3.4in}{band.eps}\caption{Band structure (in unit of $t_1$) endowed with corresponding Chern numbers by $\chi$-SDW. (a) $C=2$ Chern insulator, $U/t_1=7.15$. (b) $C=0$ Chern insulator, $U/t_1=8.0$.}
    \end{center}
    \vskip-0.5cm
    \end{figure}

To summarize, we have shown that the Hubbard model on frustrated honeycomb and bilayer triangular lattices exhibits the triple-$Q$, $\chi$-SDW order over a wide region where frustration is strongest.
Our findings provide insights into the different roles played by itinerancy, frustration, and correlation. While the existence of the vH singularity and umklapp scattering near $t_2/t_1=0.85$
produces the intervening 1$Q$-stripe and 2$Q$-spiral phases, the emergence of the $\chi$-SDW order at intermediate $U$ is a consequence of AF frustration \cite{note}. Indeed, Fig.~2 shows that direct transitions from the PM phase to the $\chi$-SDW insulator take place generically away from the special band structure point, and the phase boundary of the $C=0$ $\chi$-SDW insulator is essentially insensitive to band parameters in this region. An important corollary is that the magnetic phases of the Hubbard model are not necessarily connected adiabatically to those of the Heisenberg model with only quadratic spin exchange interactions. Recent studies of the $J_1$-$J_2$ Heisenberg model on the honeycomb lattice have not found the $\chi$-SDW phase \cite{albuquerque11,bishop12,ganesh13,zhu13,guong13}. Indeed, it is straightforward to verify using Eq.~(\ref{sdwforms}) that the 3$Q$ $\chi$-SDW, 2$Q$-spiral, and 1$Q$-stripe phases are all degenerate in the $J_1$-$J_2$ model. Additional four-site, four-spin ring exchange interaction of the form $K_2[(\vec S_1 \cdot\vec S_2)( \vec S_3 \cdot \vec S_4)+(\vec S_1\cdot \vec S_4 )(\vec S_2 \cdot \vec S_3)-(\vec S_1 \cdot \vec S_3)( \vec S_2 \cdot \vec S_4)]$ can serve to break this degeneracy and select the $\chi$-SDW as the ground state for large $U$. This is consistent with the studies of the 3D $\chi$-SDW insulator NiS$_2$ using spin models \cite{yoshida81,yoshimori81,kunitomo85,matsuura03}. For the intermediate $U$ studied here, we found no signatures of bond ordered phases and verified that the $\chi$-SDW is stable against the dimerized state in the $J_1$-$J_2$ model \cite{ganesh13}. The competition between various valence-bond order and the $\chi$-SDW can be studied using the $J_1$-$J_2$-$K_2$ model. It is hoped that our findings will further stimulate the search for topological $\chi$-SDW phases in 2D or layered hexagonal materials with both strong correlation and magnetic frustration.

This work was supported by the U.S. Department of Energy, Office of Science, Basic Energy Sciences, under Award DE-FG02-99ER45747. We thank Ying Ran, Andrej Mesaros, and Hua Chen for helpful discussions. Z.W. thanks the Aspen Center for Physics for hospitality.


\begin{thebibliography}{99}
%
\bibitem {overhauser}
 A. Overhauser, Phys. Rev.  {\bf128}, 1437 (1962).

\bibitem {luttinger}
R. Karplus, and J.M. Luttinger, Phys. Rev.  {\bf95}, 1154 (1954).

\bibitem{ye99}
Jinwu Ye, Yong Baek Kim, A. J. Millis, B. I. Shraiman, P. Majumdar, and Z. Tesanovic, Phys. Rev. Lett. {\bf83}, 3737 (1999).

\bibitem{nagaosa00}
K. Ohgushi, S. Murakami, and N. Nagaosa, Phys. Rev. B {\bf62}, R6065(R) (2000).

\bibitem{nagaosa01a}
Y. Taguchi, Y. Oohara, H. Yoshizawa, N. Nagaosa, and Y. Tokura, Science {\bf291}, 2573 (2001).

\bibitem{nagaosa01b}
R. Shindou, and N. Nagaosa, Phys. Rev. Lett. {\bf87}, 116801 (2001).

\bibitem{nagaosa10} For a recent review, see
N. Nagaosa, J. Sinova, S. Onoda, A. H. MacDonald, and N. P. Ong, Rev. Mod. Phys. {\bf82}, 1539 (2010).

\bibitem{miyadai75}
T. Miyadai, K. Takizawa, H. Nagata, H. Ito, S. Miyahara, and K. Hirakawa, J. Phys. Soc. Jpn. {\bf38}, 115 (1975).

\bibitem{kikuchi78a}
K. Kikuchi, T. Miyadai, T. Fukui, H. Ito, and K. Takizawa, J. Phys. Soc. Jpn. {\bf44}, 410 (1978).

\bibitem{kikuch78b}
K. Kikuchi, T. Miyadai, H. Itoh, and T. Fukui, J. Phys. Soc. Jpn. {\bf45}, 444 (1978).

\bibitem{matsuura03}
M. Matsuura, Y. Endoh, H. Hiraka, K. Yamada, A. S. Mishchenko, N. Nagaosa, and I. V. Solovyev, Phys. Rev. B {\bf68}, 094409 (2003).

\bibitem{endoh71}
Y. Endoh, and Y. Ishikawa, J. Phys. Soc. Jpn. {\bf30}, 1614 (1971).

\bibitem{tajima76}
K. Tajima, Y. Ishikawa, Y. Endoh, and Y. Noda, J. Phys. Soc. Jpn. {\bf41}, 1195 (1976).

\bibitem{kennedy87}
S. J. Kennedy, and T. J. Hicks, J. Phys. F {\bf17}, 1599 (1987)

\bibitem{kawarazaki90}
S. Kawarazaki, Y. Sasaki, K. Yasuda, T. Mizusaki and A Hirai, J. Phys.: Condens. Matter {\bf2}, 5747(1990)

\bibitem{yoshida81}
K. Yosida, and S. Inagaki, J. Phys. Soc. Jpn. {\bf50}, 3268 (1981).

\bibitem{yoshimori81}
A. Yoshimori, and S. Inagaki, J. Phys. Soc. Jpn. {\bf50}, 769 (1981).

\bibitem{kunitomo85}
K. Hirai, and T. Jo, J. Phys. Soc. Jpn. {\bf54}, 3567 (1985).

\bibitem{maki76}
H. Sato, and K. Maki, Prog. Theor. Phys.  {\bf55}, 319 (1976).

\bibitem{sakuma00}
A. Sakuma, J. Phys. Soc. Jpn. {\bf69}, 3072 (2000).

\bibitem{ekholm11}
M. Ekholm, and I. A. Abrikosov, Phys. Rev. B {\bf84}, 104423 (2011).

\bibitem{martin08}
I. Matrtin, and C.D. Batista, Phys. Rev. Lett. {\bf101}, 156402 (2008).

\bibitem{taoli12}
T. Li, EPL {\bf97}, 37001 (2012).

\bibitem{hayami14}

S. Hayami and Y. Motome, Phys. Rev. B {\bf90}, 060402(R), (2014).

\bibitem{chubukov12}
R. Nandkishore, L. Levitov, and A. Chubukov, Nature Phys. {\bf8}, 158 (2012).

\bibitem{wang12}
W.-S. Wang, Y.-Y. Xiang, Q.-H. Wang, F. Wang, F. Yang, and D.-H. Lee
Phys. Rev. B {\bf85}, 035414  (2012).

\bibitem{kiesel12}
M.L. Kiesel, C. Platt, W. Hanke, D.A. Abanin, and R. Thomale,
Phys. Rev. B {\bf86}, 020507(R) (2012).
\bibitem{ran14}
S. Jiang, A. Mesaros, and  Y. Ran, Phys. Rev. X {\bf4}, 031040 (2014).

\bibitem{graphenea}
K. S. Novoselov, A. K. Geim, S. V. Morozov, D. Jiang, Y. Zhang, S. V. Dubonos, I. V. Grigorieva, and A. A. Firsov, Science {\bf306}, 666 (2004).

\bibitem{grapheneb}
A. H. C. Neto, F. Guinea, N. M. R. Peres, K. S. Novoselov, and A. K. Geim, Rev. Mod. Phys. {\bf81}, 109 (2009)

 \bibitem{takada03}
K. Takada, H. Sakurai, E. Takayama-Muromachi, F. Izumi, R. A. Dilanian, and T. Sasaki, Nature (London) {\bf422}, 53 (2003).

\bibitem{singh03}
D. J. Singh, Phys. Rev. B {\bf68}, 020503(R) (2003).

\bibitem{pickett04}
K.-W. Lee, J. Kunes, and W. E. Pickett, Phys. Rev. B {\bf70}, 045104 (2004).

\bibitem{lee05}
D. Grohol, K. Matan, J.-H. Cho, S.-H. Lee, J. W. Lynn, D. G. Nocera, and Y. S. Lee, Nature Materials {\bf4}, 323 (2005).

\bibitem{shiomi12}
Y. Shiomi, M. Mochizuki, Y. Kaneko, and Y. Tokura, Phys. Rev. Lett. {\bf108}, 056601 (2012).

\bibitem{nakatsuji07}
S. Nakatsuji, H. Tonomura, K. Onuma, Y. Nambu, O. Sakai, Y. Maeno, R.T. Macaluso, and J.Y. Chan, Phys. Rev. Lett. {\bf 99}, 157203 (2007).

\bibitem{matsuda10}
M. Matsuda, M. Azuma, M. Tokunaga, Y. Shimakawa, and N. Kumada, Phys. Rev. Lett. {\bf105}, 187201 (2010).

\bibitem{singh10}
Y. Singh and P. Gegenwart, Phys. Rev. B {\bf82}, 064412 (2010).

\bibitem{li14}
Y-M Li, H-J Lun, C-Y Xiao, Y-Q Xu, L. Wu, J-H Yang, J-Y Niu, and S-C Xiang, Chem. Commun., {\bf50}, 8558 (2014).

\bibitem{kr86}
G. Kotliar, and A.E. Ruckenstein, Phys. Rev. Lett. {\bf57}, 1362 (1986).

\bibitem{li89}
T.Li, P.W$\ddot{o}$lfle and P.J. Hirschfeld, Phys. Rev. B \textbf{40}, 6817 (1989).

\bibitem{wolfle92}
R. Fr\'{e}sard, and P. W$\ddot{o}$lfle, Int. J. Mod. Phys. B \textbf{6}, 685 (1992).

\bibitem{jiang14}
K. Jiang, S. Zhou, and Z. Wang, Phys. Rev. B {\bf90}, 165135 (2014).

\bibitem{lilly90}
L. Lilly, A. Muramatsu, and W. Hanke, Phys. Rev. Lett. {\bf65}, 1380 (1990).

\bibitem{zhou14}
S. Zhou, Y.P. Wang, and Z. Wang, Phys. Rev. B {\bf89}, 195119 (2014).

\bibitem{sorella12}
S. Sorella, Y. Otsuka, and S. Yunoki, Sci. Rep. {\bf2}, 992 (2012).

\bibitem{wantanabe08}
T. Watanabe, H. Yokoyama, Y. Tanaka, and J. Inoue, Phys. Rev. B {\bf77}, 214505 (2008).

\bibitem{yoshioka09}
T. Yoshioka, A. Koga, and N. Kawakami, Phys. Rev. Lett. {\bf103} 036401 (2009).

\bibitem{lin09}
F. M. Hu, S. Q. Su, T. X. Ma, and H. Q. Lin, Phys. Rev. B {\bf80}, 014428 (2009).

\bibitem{thouless82}
D. J. Thouless, M. Kohmoto, M. P. Nightingale, and M. Den Nijs, Phys. Rev. Lett. {\bf49}, 405 (1982).

\bibitem{yao04}
Yugui Yao, Leonard Kleinman, A. H. MacDonald, Jairo Sinova, T. Jungwirth, Ding-sheng Wang,
Enge Wang, and Qian Niu, Phys. Rev. Lett. {\bf92}, 037204 (2004).

\bibitem{niu10}
Di Xiao, Ming-Che Chang, and Qian Niu, Rev. Mod. Phys \textbf{82},1959 (2010).

\bibitem{fukui05}
T. Fukui, Y. Hatsugai, and H. Suzuki, J. Phys. Soc. Jpn. \textbf{74},1674 (2005).

\bibitem{note} We verified that the $\chi$-SDW phases remain stable upon adding a small third NN hopping $t_3$, which shifts the phase bourdaries toward smaller $t_2/t_1$ in Fig.~2.

\bibitem{albuquerque11}
A. F. Albuquerque, D. Schwandt, B. Hetenyi, S. Capponi, M. Mambrini, and A. M. Lauchli, Phys. Rev. B {\bf84}, 024406 (2011).

\bibitem{bishop12}
R.F. Bishop, P. H. Y. Li, D. J. J. Farnell, and C. E. Campbell, J. Phys.: Condens. Matter {\bf24}, 236002 (2012).

\bibitem{ganesh13}
R. Ganesh, J. van den Brink, and S. Nishimoto, Phys. Rev. Lett. {\bf110}, 127203 (2013).

\bibitem{zhu13}
Z. Zhu, D. A. Huse, and S. R. White, Phys. Rev. Lett. {\bf110}, 127205 (2013).

\bibitem{guong13}
S.-S. Guong, D.N. Sheng, O. I. Motrunich, and M.P.A. Fisher, Phys. Rev. B{\bf88}, 165138 (2013).

\end{thebibliography}
\end{document}